\documentclass[conference,twocolumn,twoside]{IEEEtran}%

\usepackage{cite}

\ifCLASSINFOpdf
  \usepackage[pdftex]{graphicx}
\else
\fi
\graphicspath{{figures/}}

\usepackage[cmex10]{amsmath}
\usepackage{amssymb}

\usepackage{algorithm}
\usepackage{algpseudocode}

\usepackage[belowskip=-18pt,aboveskip=0pt,small]{caption}
\usepackage[font=footnotesize]{subfig}
\usepackage{url}

\usepackage{epstopdf}					
\usepackage{amsthm}

\theoremstyle{plain}

\newcommand\bigzero{\makebox(0,0){\text{\huge0}}}
\newcommand*{\bord}{\multicolumn{1}{c|}{}}

\usepackage{bm}

\usepackage{hyperref}

\begin{document}
%
\title{\LARGE{Reduced Complexity Optimal Detection of Binary Faster-than-Nyquist Signaling}}
%
%
%


\author{
	{Ebrahim Bedeer, Halim Yanikomeroglu, and Mohamed Hossam Ahmed\ddag}\\
	\IEEEauthorblockA{Department of Systems and Computer Engineering, Carleton University, Ottawa, ON, Canada \\ \ddag Faculty of Engineering and Applied Science, Memorial University of Newfoundland, St. John's, NL, Canada\\
		Email: \{e.bedeer, halim.yanikomeroglu\}@sce.carleton.ca, mhahmed@mun.ca}
}
\maketitle

\begin{abstract}
\boldmath In this paper, we investigate the detection problem of binary faster-than-Nyquist (FTN) signaling and propose a novel sequence estimation technique that exploits its special structure. In particular, the proposed sequence estimation technique is based on sphere decoding (SD) and exploits the  following two characteristics about the FTN detection problem: 1) the correlation between the noise samples after the receiver matched filter,  and 2) the  structure of the intersymbol interference (ISI) matrix. Simulation results show that the proposed SD-based sequence estimation (SDSE) achieves the optimal performance of the maximum likelihood sequence estimation (MLSE) at reduced computational complexity. 
This paper demonstrates that  FTN signaling has the great potential of  increasing the data rate and spectral efficiency substantially, when compared to Nyquist signaling, for the same bit-error-rate (BER) and signal-to-noise ratio (SNR).
\end{abstract}




\section{Introduction} \label{sec:into}

Recently, faster-than-Nyquist (FTN) signaling \cite{anderson2013faster} re-attracted the attention of the research community as a promising transmission technique that is capable of improving the spectral efficiency in both wireless and wired communication systems.
FTN signaling refers to the transmission of pulses beyond the Nyquist limit. Nyquist limit is the signaling  threshold 
beyond which intersymbol interference (ISI) between the received pulses is unavoidable. More specifically, Nyquist showed that signaling at rates greater than $1/T$ of $T$-orthogonal pulses, i.e., pulses that are orthogonal to an $nT$ shift of themselves for nonzero integer $n$, results in  ISI at the samples of receiver's matched filter output \cite{nyquist1928certain}.

On the contrary of what is widely known that the \textit{term} FTN was coined by J. Mazo in \cite{mazo1975faster}, it seems that the FTN signaling term has been around even earlier as evidenced by the works of Lucky \cite{lucky1970decision} and Salz \cite{salz1973optimum}. Moreover, the concept of violating the Nyquist limit seems to appear even earlier as shown in the work of Saltzberg in \cite{saltzberg1968intersymbol} (dual of Mazo's formulation, where the data rate was maintained the same while decreasing the transmit pulse bandwidth). The recognition of Mazo's work comes from the fact that 
Mazo was the first to prove that FTN signaling does not affect the minimum distance of uncoded binary transmission, and hence the asymptotic error probability, as long as the signaling rate is below a certain limit, later became known as Mazo limit. In particular, Mazo investigated the case of binary cardinal sine (i.e., sinc) pulse transmission and showed that binary sinc pulses can be accelerated to a signaling rate $1/(\tau T)$. Such an acceleration is while the minimum distance remains the same as long as $\tau \in \left[0.802, 1\right]$, despite the ISI between the received pulses. This means that up to $\frac{1}{0.802} - 1 \simeq 25 \%$ more bits can be transmitted in the same bandwidth at the same energy per bit without degrading the bit error rate (BER).

The uncoded transmission of FTN signaling was viewed as a trellis coding method in \cite{prlja2008receivers} and a truncated state Viterbi algorithm (VA) was proposed as an efficient detection scheme to reduce the computational complexity of full states VA algorithms. Reduced state Bahl-Cocke-Jelinek-Raviv (BCJR) algorithms were proposed to balance performance and computational complexity in \cite{anderson2009new}. 
The works in \cite{prlja2008receivers, anderson2009new} are still complex and are more suitable for severe ISI scenarios, i.e., very tight packing/acceleration of pulses in time, where other conventional equalization techniques fail to give satisfactory performance.
For low ISI scenarios,  the authors in \cite{bedeer2016very} identified an operating region (that depends on the pulse shape, its roll-off factor, and the  time packing/acceleration parameter), where perfect reconstruction of FTN signaling is guaranteed for noise-free transmission. For noisy transmission, they proposed two novel algorithms (with the lowest complexity reported in the literature so far) to detect FTN signaling on a symbol-by-symbol basis.
In   \cite{ishihara2016frequency}, the authors extended the  frequency domain equalizer (FDE) in \cite{sugiura2013frequency}   to produce soft-decision of the estimated data symbols  while considering the correlated noise samples after the receiver matched filter.  The authors in \cite{bedeer2016low} proposed a novel algorithm to detect any high-order quadrature amplitude modulation FTN signaling, in polynomial time complexity, and  efficiently works for moderate values of the time packing/acceleration parameter.

In this paper, we explore novel reduced complexity sphere decoding (SD)-based  sequence estimation technique to detect binary FTN signaling. 
The main contributions of this paper are summarized as follows:
\begin{itemize}
	\item In general, for ISI channels, the noise samples at the receiver matched filter output are non-white. Hence, we propose a generalized SD-based sequence estimation (SDSE) that can handle such instances of colored noise. This is mainly achieved by incorporating the whitening filter into the standard SD algorithm. We then apply the proposed SDSE to the detection  of binary FTN signaling that has colored noise samples after the receiver matched filter.
	\item Simulation results show that the  proposed SDSE achieves the BER performance of maximum likelihood sequence estimation (MLSE). 
	Additionally, results show that FTN can significantly increase the data rate, when compared to Nyquist signaling, for the same BER and signal-to-noise ratio (SNR).
\end{itemize}

 

The remainder of this paper is organized as follows. Section \ref{sec:model} presents the system model of the FTN signaling and formulates the MLSE problem to detect binary FTN signaling. The proposed SDSE is discussed in Section \ref{sec:SD}.
Section \ref{sec:results} provides the simulation results of our proposed sequence estimation techniques, and finally the paper is concluded in Section \ref{sec:conc}.

Throughout the paper we use bold-faced upper case letters, e.g., $\bm{X}$, to denote matrices, bold-faced lower case letters, e.g., $\bm{x}$, to denote column vectors, and light-faced italics letters, e.g., $x$, to denote scalars. $X_{i,j}$ denotes the element in the $i$th row and the $j$th column of the matrix $\bm{X}$, and $x_i$ denotes the $i$th element of vector $\bm{x}$. 
$[.]^{\rm{T}}$ denotes the transpose operator, 
$\bm{I}$ is the identity matrix, 
$\mathbb{E}(.)$ is the expectation operator, $\Vert . \Vert_p$ is the $p$-norm, and $\mathcal{N}(.,.)$ represents the Gaussian distribution. $\lceil.\rceil$ denotes rounding to the nearest larger integer within a given set and $\lfloor.\rfloor$  rounding to the nearest smaller integer in a given set.

\section{System Model and MLSE Problem Formulation} \label{sec:model} 

\begin{figure}[!t]
	\centering
	\includegraphics[width=0.50\textwidth]{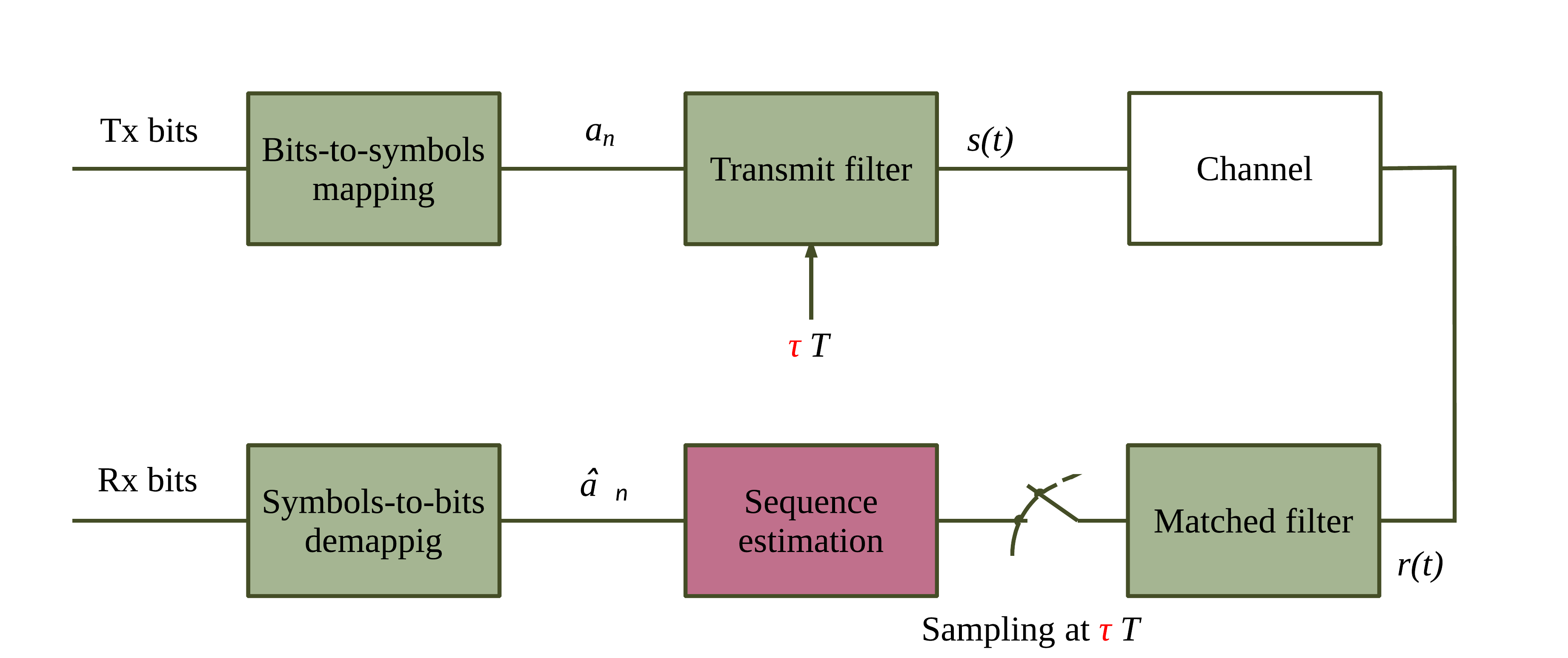}
	\caption{Block diagram of FTN signaling.}\label{fig:block_diagram}
\end{figure}

Fig. \ref{fig:block_diagram} shows a  block diagram of a communication system employing FTN signaling. Data bits to be transmitted are gray mapped to data symbols through the bits-to-symbols mapping block. Data symbols are transmitted, through the transmit filter block, faster than Nyquist signaling, i.e., every $\tau T$, where $0 < \tau \leq 1$ is the time packing/acceleration parameter and $T$ is the symbol duration. A possible receiver structure is shown in Fig. \ref{fig:block_diagram}, where the received signal is passed through a filter matched to the transmit filter followed by a sampler. Since the transmission rate of the transmit pulses carrying the data symbols intentionally violate the Nyquist criterion, ISI exists between the received samples. Accordingly, sequence estimation techniques are needed to remove the ISI and to estimate the transmitted data symbols. 
The estimated data symbols are finally gray demapped to the estimated received bits.

The transmitted signal $s(t)$ of the FTN signaling shown in Fig. \ref{fig:block_diagram} can be written in the form
\begin{eqnarray}
s(t) = \sqrt{E_s} \: \sum\nolimits_{n = 1}^{N} a_n \: p(t - n \tau T), \qquad 0 < \tau \leq 1,
\end{eqnarray} 
where $N$ is the total number of transmit data symbols, $a_n, \: n= 1, \hdots, N,$ is the independent and identically distributed  data symbols, $E_s$ is the data symbol  energy, $p(t)$ is a unit-energy pulse, i.e., $\int\nolimits_{-\infty}^{\infty} \vert p(t) \vert^2 dt = 1$,  and $1/(\tau T)$ is the signaling rate. 
The received FTN signal in case of additive white Gaussian noise (AWGN) channel is written as
\begin{eqnarray}
y(t) = s(t) + \bar{w}(t), 
\end{eqnarray}
where $\bar{w}(t)$ is a zero mean complex valued Gaussian random variable with power spectral density  $\sigma^2$. A possible receiver architecture for FTN signaling is to use a filter matched to $p(t)$; thus the received signal after the matched filter can be written as
\begin{eqnarray}
y(t) = \sqrt{E_s} \: \sum\nolimits_{n = 1}^{N} a_n g(t - n \tau T) + w(t),
\end{eqnarray}
where $g(t) = \int\nolimits p(x) p(x - t) dx$ and $w(t) = \int\nolimits \bar{w}(x) p(x - t) dx$. The noise of FTN signaling after the  matched filter, i.e., $w(t)$, is zero-mean with a covariance matrix given by
\begin{eqnarray}
{\rm{Cov}}(w(t_1),w(t_2)) \hspace{5cm}
\end{eqnarray} \vspace{-20pt}
\begin{eqnarray}
& = &  \mathbb{E}\Big[\big[w(t_1) - \mathbb{E}(w(t_1))\big] \big[w(t_2) - \mathbb{E}(w(t_2))\big]\Big] \nonumber \\
& = & \mathbb{E}\bigg[ \int\nolimits_{-\infty}^{\infty} \bar{w}(t_1) p(t_1 - n \tau T) \: dt_1   \nonumber \\ && \hspace{3cm} \int\nolimits_{-\infty}^{\infty} \bar{w}(t_2) p(t_2 - n' \tau T) \: dt_2 \bigg] \nonumber \\
& = & \int\nolimits_{-\infty}^{\infty} \int\nolimits_{-\infty}^{\infty} \mathbb{E} \big[\bar{w}(t_1) \bar{w}(t_2)] \: p(t_1 - n \tau T) \nonumber \\ && \hspace{4cm} p(t_2 - n' \tau T) \: dt_1 dt_2.
\end{eqnarray}
Since $\mathbb{E} \big[\bar{w}(t_1) \bar{w}(t_2)] = \sigma^2 \delta(t_1 - t_2)$, the covariance matrix of the noise $w(t)$ after the matched filter is written as
\begin{IEEEeqnarray}{rcl}\label{eq:cov}
{\rm{Cov}}(w(t_1),w(t_2)) & {}={} & \sigma^2 \int\nolimits_{-\infty}^{\infty} p(t_1 - n \tau T)  p(t_1 - n' \tau T) \: dt_1 \nonumber \\
& = & \sigma^2 \: G_{n,n'},
\end{IEEEeqnarray}
where $G_{n,n'} = g((n - n') \tau T)$ represents the ISI between data symbols $n$ and $n'$.
As can be seen, the noise covariance matrix is not diagonal, and hence, the elements of the noise vector $\bm{w}$, after the matched filter, are dependent.
Assuming perfect timing synchronization between the transmitter and the receiver, the received FTN signal $y(t)$ is sampled every $\tau T$ and  the $k$th received sample can be expressed as
\begin{IEEEeqnarray}{rcl}
y_k &{}={}& y(k \tau T) \nonumber \\
	&{}={}& \sqrt{E_s} \sum\nolimits_{n = 1}^{N} a_n g(k \tau T - n \tau T) + w(k \tau T) \nonumber \\
	&{}={}& \sqrt{E_s} \: a_k \: g(0) \nonumber \\ \hfill & & +  \sqrt{E_s} \: \sum\nolimits_{n = 1, \: n \ne k}^{N} a_n \: g((k - n) \tau T) + w(k \tau T).
\end{IEEEeqnarray} 
Finally, the received FTN signal after the matched filter and sampler can be written in a vector form as 
\begin{eqnarray} \label{eq:vector}
\bm{y} = \bm{G} \: \bm{a} + \bm{w},
\end{eqnarray}
where $\bm{a}$ is the transmitted data symbols, $\bm{w} \sim \mathcal{N}(0, \sigma^2 \bm{G})$ is the Gaussian noise samples with zero-mean and covariance matrix $\sigma^2 \bm{G}$, and the ISI matrix $\bm{G}$  is given as 
\begin{IEEEeqnarray}{rcl}\label{eq:ISI}
\bm{G} & = &  \begin{bmatrix}
G_{1,1} & G_{1,2}    & \hdots  & G_{1,L} & 0 &  0 & 0 \\ 
G_{1,2} & G_{1,1}    & \hdots & \hdots  & G_{1,L} &  0 & 0  \\ 
\ddots & \ddots   & \ddots  & \ddots  & \ddots &  \ddots &     \ddots\\ 
\hdots & \hdots   & G_{1,2}  & G_{1,1} & G_{1,2}   & \hdots &     \hdots\\ 
\ddots & \ddots  & \ddots & \ddots & \ddots &  \ddots &     \ddots\\ 
0 & 0 &  G_{1,L} & \hdots & \hdots  & G_{1,1} & G_{1,2}\\ 
0 & 0  & 0 & G_{1,L} & \hdots  & G_{1,2} & G_{1,1}
\end{bmatrix}. \IEEEeqnarraynumspace 
\end{IEEEeqnarray}
The ISI length of FTN signaling is theoretically infinite; however, it is practical to truncate it to a certain length $L$ that depends on the transmit and matched filters and the value of $\tau$ \cite{prlja2008receivers, anderson2009new}. 
Hence,  for a large value of $N$ (i.e., long block transmission) and high/moderate values of $\tau$ (non-severe ISI), the ISI matrix can be  sparse due to the fact that  a given symbol is affected only by $L \ll N$ adjacent symbols.
One can show that the interference matrix $\bm{G}$ is invertible (the proof is omitted due to space limitations); hence the received noisy data symbol vector $\bm{y}$ in \eqref{eq:vector} can be rewritten as
\begin{eqnarray} \label{eq:vector_z}
	\bm{G}^{-1} \: \bm{y} &=& \bm{a} + \bm{G}^{-1} \: \bm{w} \nonumber \\
	\bm{z} &=& \bm{a} + \bm{\eta},
\end{eqnarray}
where $\bm{z} = \bm{G}^{-1} \: \bm{y}$ and $\bm{\eta} = \bm{G}^{-1} \: \bm{w}$. 
The FTN detection problem can be interpreted as follows. Given the received samples $\bm{z}$ (or $\bm{y}$), we want to find an estimated data symbol vector $\hat{\bm{a}}$, to the transmit data symbol vector $\bm{a}$, such that the probability of error is minimized. After some mathematical manipulations, the  MLSE problem to detect binary FTN signaling can be expressed as

\begin{eqnarray}\label{eq:opML}
\mathcal{OP}_{\rm{MLSE}}: &&  \underset{\bm{a}}{\min} \quad (\bm{z} - \bm{a})^{\rm{T}} \: \bm{G} \: (\bm{z} - \bm{a}) \nonumber \\
	& & {\textup{subject to}} \qquad \bm{a} \in \{1 , -1\}^N.
\end{eqnarray} 
The $\mathcal{OP}_{\rm{MLSE}}$ problem can be solved by a brute force search; however, such extensive computational complexity hinders its practical implementations. 
\section{Binary FTN signaling Detection using the Proposed SDSE} \label{sec:SD}
In this section, we discuss the proposed SDSE that exploits the following two characteristics of the FTN problem: 1) the fact that the noise samples after the matched filter are non-white, and 2) the structure of the ISI matrix (i.e., the possible sparsity). 
These two features allow the proposed SDSE to reach the MLSE performance at reduced complexity.

\subsection{Main Idea of Standard SD Algorithm}\label{sec:standard_SD}
The SD algorithm was originally proposed in \cite{fincke1985improved} to optimally solve the following integer least-square problem
\begin{eqnarray}\label{eq:standard_SD}
\hat{\bm{\overline{a}}} &=& \arg \: \underset{\bm{\overline{a}} \: \in \: \mathcal{D}}{\min} \quad \Vert \bm{\overline{y}} - \bm{\overline{G}} \bm{\overline{a}} \Vert_2^2,
\end{eqnarray}
where $\mathcal{D}$ is a subset of the integer lattice $\mathbb{Z}^N$. Such a least-square problem appears in communication systems to detect the data symbols vector $\bm{\overline{a}}$ from the received vector $\bm{\overline{y}}$ given as
\begin{eqnarray}\label{eq:detect_white_noise}
\bm{\overline{y}}   &=& \bm{\overline{G}} \bm{\overline{a}} + \bm{\overline{\eta}},
\end{eqnarray}
where $\bm{\overline{\eta}} \sim \mathcal{N}(o,\sigma^2 \bm{I})$ is a zero-mean Gaussian random variable with covariance matrix $\sigma^2 \bm{I}$ and $\bm{\overline{G}}$ is the channel coefficient matrix. In other words, the SD algorithm was originally proposed to find the closest point in a lattice, which is equivalent to digital symbol detection in the presence of white noise.


The main idea of sphere decoding is to search over those lattice points that lie inside a hypersphere of radius $d$ around the received vector $\bm{\overline{y}}$. Hence, the search space and the required computations are reduced when compared to the brute force search required for the proper operator of MLSE. Clearly, the closest lattice point inside the hypersphere is also the closest lattice point in the whole lattice, and hence, the solution from the SD algorithm is the optimal solution of MLSE. 

It is important to carefully choose the radius $d$ to avoid the two extreme cases, i.e., 1) if $d$ is chosen to be very large, then there will be too many lattice points inside the hypersphere to evaluate and 2) if $d$ is chosen to be too small, then there might be no points inside the hypersphere and no solution can be reached. Generally, it is efficient to choose the initial radius to be the distance between the received sample $\bm{\overline{y}}$ and the zero-forcing solution $\bm{\overline{a}}_{\rm{ZF}}$ \cite{hassibi2005sphere}, i.e.,   $d = \Vert \bm{\overline{y}} - \bm{\overline{G}} \bm{\overline{a}}_{\rm{ZF}} \Vert^2_2$.

A lattice point lies inside a hypersphere centered at $\bm{\overline{y}}$ of radius $\overline{d}$ if and only if its radius $\overline{d}$ is chosen such that 
\begin{eqnarray}\label{eq:radius_white}
\overline{d}^2 &\geq& \Vert \bm{\overline{y}} - \bm{\overline{G}} \bm{\overline{a}} \Vert_2^2.
\end{eqnarray}
In order to divide the problem into problems of lower dimensions, we apply $\bm{Q} \bm{R}$ factorization of the channel coefficient matrix $\bm{\overline{G}}$ as $\bm{\overline{G}} = \bm{\overline{Q}} \bm{\overline{R}}$, and write \eqref{eq:radius_white} as 
\begin{eqnarray}\label{eq:radius_white_all}
\overline{d}^2 &\geq& \Vert \bm{\overline{z}} - \bm{\overline{R}} \bm{\overline{a}} \Vert_2^2, \nonumber \\
\overline{d}^2 &\geq& \sum\nolimits_{i = 1}^{N} \left( \overline{z}_i - \sum\nolimits_{j = i}^{N} \overline{R}_{i,j} \overline{a}_j \right)^2, 
\end{eqnarray}
where $\bm{\overline{z}} = \bm{\overline{Q}}^{\rm{T}} \bm{\overline{y}}$. The property of the upper triangular $\bm{\overline{R}}$ is useful in the sense that the right-hand side (RHS) of the above inequality can be expanded as
\begin{IEEEeqnarray}{rcl}\label{eq:SD_terms}
\overline{d}^2 &{}\geq{}& (\overline{z}_N  - \overline{R}_{N,N} \: \overline{a}_N)^2 \nonumber \\ & & + (\overline{z}_{N-1} - \overline{R}_{N-1,N-1} \: \overline{a}_{N-1} - \overline{R}_{N-1,N} \: \overline{a}_{N})^2 \nonumber \\ & & + \hdots.
\end{IEEEeqnarray}
As can be seen, the first term of the RHS of the above inequality depends only on $\overline{a}_N$, the second terms depends on $\{\overline{a}_N, \overline{a}_{N - 1}\}$, and so on. As such, a necessary condition for the received sample $\bm{\overline{z}}$ (or $\bm{\overline{y}}$) to lie inside the hypersphere of radius $\overline{d}$ is that 
\begin{eqnarray}\label{eq:start_point}
\overline{d}^2 &\geq& (\overline{z}_N  - \overline{R}_{N,N} \: \overline{a}_N)^2,
\end{eqnarray}
which is equivalent to the $N$th data symbol $\overline{a}_N$ belongs to the following interval
\begin{eqnarray}\label{eq:interval_SD}
\left\lceil\frac{\overline{z}_N - \overline{d}}{\overline{R}_{N,N}} \right\rceil  \leq \: \overline{a}_N \: \leq  \left\lfloor \frac{\overline{z}_N + \overline{d}}{\overline{R}_{N,N}} \right\rfloor,
\end{eqnarray}
It is clear that \eqref{eq:interval_SD} does not guarantee that the whole vector ${\bm{\overline{a}}}$ lies inside the hypersphere of radius $\overline{d}$. To find all the lattice points that lie inside the hypersphere centered at $\bm{\overline{z}}$ (or $\bm{\overline{y}}$) with a radius $\overline{d}$, let us define $\overline{d}^2_{N-1} = \overline{d}^2 - (\overline{z}_N  - \overline{R}_{N,N} \: \overline{a}_N)^2$, then from \eqref{eq:SD_terms}, we can impose the following constraints on the $(N-1)$th symbol $\overline{a}_{N-1}$ 
\begin{eqnarray}\label{eq:interval_SD_N_1}
\overline{a}_{N - 1} &\geq& \left\lceil\frac{\overline{z}_{N-1} - \overline{R}_{N-1,N} \: \overline{a}_{N} - \overline{d}_{N-1}}{\overline{R}_{N,N}} \right\rceil,   \\
\overline{a}_{N - 1}  &\leq&  \left\lfloor \frac{\overline{z}_{N-1} - \overline{R}_{N-1,N} \: \overline{a}_{N} + \overline{d}_{N-1}}{\overline{R}_{N,N}} \right\rfloor.\label{eq:interval_SD_N_11}
\end{eqnarray}
One can continue the process to find the constraints on the intervals for other symbols $\overline{a}_{N - 2}, \hdots, \overline{a}_{1}$ until we found all the lattice points ${\bm{\overline{a}}}$ inside the hypersphere.

\subsection{Proposed SDSE for Binary FTN Detection}\label{sec:modified_SD}
As discussed earlier, one of the main characteristics of the FTN detection problem in \eqref{eq:vector_z} is that the noise samples are no longer white at the sampling intervals $\tau T$ due to the non-orthogonality between the transmit pulses. In this subsection, we exploit such observation  to propose a reduced complexity SDSE\footnote{The proposed SDSE provides hard-decisions on the estimated data symbols. In our future works, we will extend the proposed SDSE to provide soft-decisions, and hence, enables its use with channel coding.} that achieves the MLSE performance. In particular, this is achieved through incorporating the whitening filter into the SD algorithm.
Before introducing the proposed SDSE, let us reformulate the objective function of the MLSE optimization problem $\mathcal{OP}_{\rm{MLSE}}$  in \eqref{eq:opML} as 
\begin{eqnarray}
(\bm{z} - \bm{a})^{\rm{T}} \bm{\bm{R}}^{\rm{T}} \bm{\bm{R}} (\bm{z} - \bm{a}) 
& = &  \Vert \bm{\bm{R}}(\bm{z} - \bm{a}) \Vert^2_2,
\end{eqnarray}
where $\bm{R}$ is an upper triangular matrix and $\bm{G} = \bm{R}^{\rm{T}} \bm{R}$ is the Cholesky decomposition of $\bm{G}$. Hence, the optimal binary detection of FTN signaling problem can be re-expressed as
\begin{eqnarray}\label{eq:FTN_detection_SD}
&&\underset{\bm{a}}{\min} \quad \Vert \bm{\bm{R}}(\bm{z} - \bm{a}) \Vert^2_2,  \nonumber \\
	& & {\textup{subject to}} \qquad \bm{a} \in \{1 , -1\}^N.
\end{eqnarray}
It is important to note that the Cholesky decomposition of $\bm{G}$ as $\bm{G} = \bm{R}^{\rm{T}} \bm{R}$ results in the following interesting observation. The vector $\bm{\bm{R}}(\bm{z} - \bm{a})$ which is equivalent to ${\bm{R}} \: \bm{\eta} = \bm{R} \: \bm{G}^{-1} \: \bm{w}$ is the same as  the whitening matched filter in \cite{forney1972maximum}. This can be verified from finding the covariance matrix of the noise vector $\bm{R} \: \bm{G}^{-1} \: \bm{w}$ which will be $\sigma^2 \bm{I}$. Accordingly, the new detection problem in \eqref{eq:FTN_detection_SD} is equivalent to having the the received sample $\bm{z}$ perturbed by a noise in a hypersphere.

If the ISI matrix $\bm{G}$ is sparse (this is especially true for transmission of long blocks of data symbols, i.e., high values of $N$, in a non-severe and moderate interference scenarios, i.e., high and medium values of $\tau$), then its Cholesky factorization is often sparse as well \cite[Chapter 6]{vandenberghe2011applied}. This means that $\bm{R}$ is an upper triangular matrix, where it contains at most $L$ non-zero elements in the first $N - L$ rows, and $L-1, L-2, \hdots, 1$ non-zero elements for the last $L - 1, L-2, \hdots, 1$ rows, respectively. In other words, the upper triangular matrix $\bm{R}$ can be viewed as\vspace{-7pt}
\begin{eqnarray}
\bm{R} = \hspace{7cm} 
\end{eqnarray}\vspace{-12pt}
{\small
\begin{eqnarray}
\left[\begin{array}{cccccccc}
R_{1,1}    & R_{1,2}           & \hdots    & R_{1,L} & 0 & \hdots & \hdots \\ \cline{1-1}
\bord & R_{2,2}           & \hdots     & \hdots & R_{2,L + 1} & 0  &\hdots \\ \cline{2-2}
&  \bord      & \ddots & \ddots & \ddots & \ddots \\ \cline{3-3}
&      \bigzero           & \bord & R_{\ell,\ell} &  \hdots & R_{\ell,\ell + L} & \ddots\\ \cline{4-4}
&                 &  & \bord & \ddots &  \ddots & \ddots\\ \cline{5-5}
&          &         & & \bord & R_{N-1,N-1}  & R_{N-1,N}\\ \cline{6-6}
&          &       &    & & \bord & R_{N,N}\\ \cline{7-7}
\end{array}\right]. \nonumber
\end{eqnarray}}
Hence, an equivalent MLSE optimization problem to detect binary FTN signaling to the one in \eqref{eq:FTN_detection_SD} can be written as
\begin{eqnarray}\label{eq:halim}
\mathcal{OP}_{\rm{SD}}: & & \underset{\bm{a}}{\min} \: \sum\nolimits_{i = 1}^{N} \sum\nolimits_{j = i}^{\min(L + i - 1, N)} \left( R_{i,j} (z_j - a_j) \right)^2, \nonumber \\
	& & {\textup{subject to}} \: \bm{a} \in \{1 , -1\}^N.
\end{eqnarray}
It is worthy emphasizing  that the inner summation over $j$ is only for $L$ elements (for the first $N - L$ rows in $\bm{R}$) or less (for the last $L - 1$ rows in $\bm{R}$). This represents further reduction in the  complexity and memory requirements when compared to the equivalent objective of the standard SD algorithm in \eqref{eq:radius_white_all}, where the summations are evaluated over $N$ elements.

According to the objective of $\mathcal{OP}_{\rm{SD}}$ in \eqref{eq:halim} and in order to guarantee that the lattice point  lies inside a hypersphere centered at $\bm{z}$ (or $\bm{y}$) of radius $d$,  the initial radius of the proposed SDSE is chosen such that
\begin{eqnarray}\label{eq:radius}
d^2 &\geq& \sum\nolimits_{i = 1}^{N} \sum\nolimits_{j = i}^{\min(L + i - 1, N)} \left( R_{i,j} (z_j - a_j) \right)^2 \nonumber \\
&\geq& \left( R_{N,N} (z_N - a_N) \right)^2 + \big( R_{N-1,N-1} (z_{N-1} - a_{N-1}) \nonumber \\ & &  + R_{N-1,N} (z_{N} - a_{N}) \big)^2 + \hdots \:.
\end{eqnarray}

As can be seen, the first term in \eqref{eq:radius} depends only on the last data symbol $a_N$, the second term depends on the last two data symbols $\{a_N, a_{N-1}\}$, and so on, until the last term that would depend on the whole vector of the data symbols $\bm{a}$.
A necessary condition for the $N$th data symbol $a_N$ to lie within the hypersphere is
\begin{eqnarray}
d^2 &\geq& \left( R_{N,N} (z_N - a_N) \right)^2,
\end{eqnarray}
which is equivalent to the $N$th data symbol lies in the following interval
\begin{eqnarray}\label{eq:last_symbol}
\left\lceil z_N - \frac{d}{R_{N,N}} \right\rceil  \leq \: a_N \: \leq  \left\lfloor z_N + \frac{d}{R_{N,N}} \right\rfloor.
\end{eqnarray}
Similarly, one can find the interval which the ($N-1$)th data symbol belongs to as
\begin{eqnarray}\label{eq:N_1_symbol}
a_{N - 1} &\geq& \left\lceil z_{N - 1} - \frac{\hat{d} - R_{N-1,N} (z_N - a_N)}{R_{N - 1,N - 1}} \right\rceil, \\
a_{N - 1}  &\leq& \left\lfloor z_{N - 1} + \frac{\hat{d} + R_{N-1,N} (z_N - a_N)}{R_{N - 1,N - 1}} \right\rfloor, \label{eq:N_1_symbol1}
\end{eqnarray}
where 
\begin{eqnarray}
\hat{d}^2 = d^2 - \left( R_{N,N} (z_N - a_N) \right)^2. 
\end{eqnarray}
This process continues to find the interval boundaries for all other symbols $a_{N-2}, a_{N-3}, \hdots, a_1$. It is worthy to emphasize that exploiting the information in the noise covariance matrix, leads to different radii (\eqref{eq:radius} versus \eqref{eq:SD_terms}) and interval limits (\eqref{eq:last_symbol},  \eqref{eq:N_1_symbol}, and \eqref{eq:N_1_symbol1} versus \eqref{eq:interval_SD}, \eqref{eq:interval_SD_N_1}, and \eqref{eq:interval_SD_N_11}, respectively) of the proposed SDSE when compared to a standard SD algorithm. 

The tree search of the proposed SDSE can be explained as follows. All possible values of the $N$th transmit symbol $a_{N}$ 
can be determined from the interval boundary as in \eqref{eq:last_symbol} and each one of these values will represent a node at the $N$th level. 
We pick a node  to branch, and we  find a possible node in the $(N-1)$th level, i.e., one of the values of the $(N-1)$th data symbol $a_{N-1}$ according to its interval in \eqref{eq:N_1_symbol} and \eqref{eq:N_1_symbol1}. The branching process continues by selecting a node in each level until reaching first level. Then, the distance of this new found vector $\bm{a}$ to the received sample $\bm{z}$ is calculated and the distance $d$ in \eqref{eq:radius} is updated (i.e., reduced). For each level, we branch only nodes where further branching will not increase that distance. Everytime we reach the first level, we update the distance $d$, and the process continues until all nodes are either branched or pruned.

The proposed SDSE is formally expressed as follows:

{\textit{Algorithm 1:} Proposed SDSE}
\begin{enumerate}
		\item \textbf{Input:} Pulse shape $p(t)$, $\bm{G}$, and received samples $\bm{y}$.
		\item Perform Cholesky decomposition of $\bm{G}$ as $\bm{G} = \bm{R}^{\rm{T}} \bm{R}$.
		\item Set initial radius as $d^2 \geq \left( R_{N,N} (z_N - a_{\rm{ZF}_N}) \right)^2 $.
		\item Find all $a_N$ in the interval in \eqref{eq:last_symbol}.
		\item For a given value of $a_N$, find $a_{N-1}$ at the $(N-1)$th level of the tree  according to its interval in \eqref{eq:N_1_symbol} and \eqref{eq:N_1_symbol1}.
		\item For a given value of $a_{N-1}$, continue the branching process until reaching a symbol $a_1$ at the first level. If for a given node at any level the new distance exceeds the radius, then do not further branch at this node.
		\item Update radius with the obtained candidate data symbol.
		\item Repeat steps 5, 6, and 7 until all the nodes are either branched or pruned.
		\item \textbf{Output:} Estimated data symbols. 
\end{enumerate}


\subsection{Complexity Discussion}
As discussed earlier, the proposed SDSE requires factorization of a sparse matrix (in case of non-severe ISI), while a standard SD algorithm requires factorization of a dense matrix. The number of required arithmetic operations to perform the factorization in a standard SD algorithm  is in the order of $N^3$ \cite{vandenberghe2011applied}, while for sparse factorization this number in general  depends on the number of non-zero elements. For the positive definite ISI matrix $\bm{G}$, the number of required arithmetic operations is in the order of $N^{1.5}$ \cite{vandenberghe2011applied}. 

Another source of complexity reduction for the proposed SDSE can be seen in the steps of calculating/updating the radius. In particular, we do the summation over only $L \ll N$ terms (compared to $N$ terms for the standard SD algorithm), which represents a reduction in computational complexity and memory requirements.

\section{Simulation Results} \label{sec:results}
In this section, we evaluate the performance of the proposed SDSE in detecting binary FTN signaling. We employ an rRC filter with roll-off factors $\beta = 0.3$ and  $0.5$, and we consider data symbols to be drawn from the constellation of  binary phase shift keying (BPSK). The spectral efficiency is calculated as $\frac{\log_2 M}{(1 + \beta) \: \tau}$, where $M = 2$ is the BPSK constellation size.


\subsection{Performance of the Proposed SDSE}
\begin{figure}[!t]
	\centering
	\includegraphics[width=0.50\textwidth]{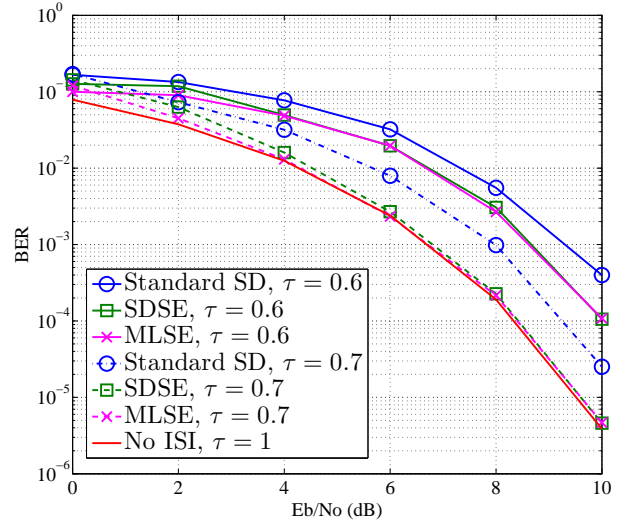}
	\caption{BER performance of binary FTN detection versus $\frac{E_b}{N_o}$ using the standard SD-based and proposed SDSEs  at $\beta = 0.3$ and $\tau = 0.6$ and $0.7$.}\label{fig:SD_standard_vs_proposed_alpha_03_tau_06_07}
\end{figure}
Fig. \ref{fig:SD_standard_vs_proposed_alpha_03_tau_06_07} depicts the BER of binary FTN as a function of $\frac{E_b}{N_o}$ for the standard SD algorithm, proposed SDSE, and the MLSE   for $\beta = 0.3$ and  $\tau = 0.6$ and 0.7. As can be seen in Fig. \ref{fig:SD_standard_vs_proposed_alpha_03_tau_06_07}, the proposed SDSE that exploits the information in the noise samples covariance matrix outperforms the standard SD algorithm and achieves the MLSE performance of $\mathcal{OP}_{\rm{MLSE}}$. 
Further and as expected, increasing the value of $\tau$ improves the BER performance of both standard and proposed SDSE. Fig. \ref{fig:SD_standard_vs_proposed_alpha_03_tau_06_07} reveals that the proposed SDSE can achieve $\frac{1}{0.7} - 1= 42.86 \%$  increase in the transmission rate at $\beta = 0.3$ and $\tau = 0.7$ without harming the BER when compared to the Nyquist signaling (i.e., no ISI case).


\begin{figure}[!t]
	\centering
	\includegraphics[width=0.50\textwidth]{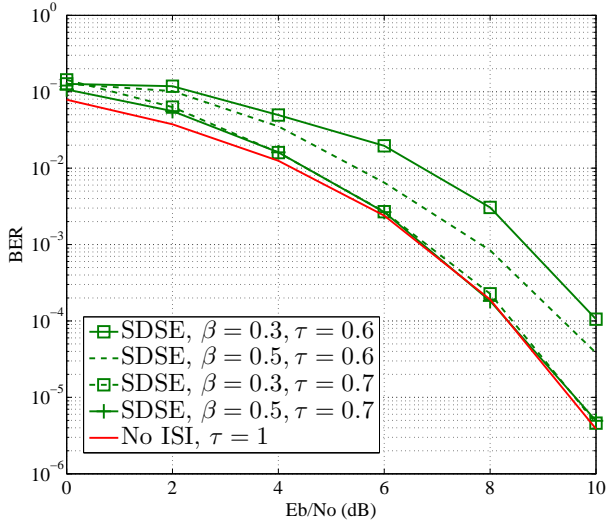}
	\caption{BER performance of binary FTN detection versus $\frac{E_b}{N_o}$ using the proposed SDSE   at $\beta = 0.3$ and $0.5$, and $\tau = 0.6$ and $0.7$.}\label{fig:SD_proposed}
\end{figure}
Fig. \ref{fig:SD_proposed} plots the BER of binary FTN detection as a function of $\frac{E_b}{N_o}$ for the proposed SDSE  for roll-off factors of $\beta = 0.3$ and $0.5$ and $\tau = 0.6$ and $0.7$. One can notice from Fig. \ref{fig:SD_proposed} that increasing the value of $\beta$ improves the BER performance of FTN signaling. This is expected and can be explained as follows: as the value of $\beta$ increases, the rRC pulse amplitude decays more rapidly in time domain, and hence, the ISI between adjacent symbols are reduced. One can infer from Fig. \ref{fig:SD_proposed} that for $\tau = 0.7$ and $\beta = 0.3$ or $0.5$, the proposed SDSE can enable the transmission of $\frac{1}{0.7} - 1 = 42.86 \%$ more bits, when compared to Nyquist signaling, in the same bandwidth at the same energy per bit without degrading the BER. Further, for $\tau = 0.6$, $\frac{1}{0.6} - 1 = 66.67 \%$ more bits can be transmitted compared to the Nyquist signaling at $\beta = 0.3$ and $0.5$, at the expense of 1.5 dB and 1 dB, respectively, at BER = $10^{-4}$.

\begin{figure}[!t]
	\centering
	\includegraphics[width=0.50\textwidth]{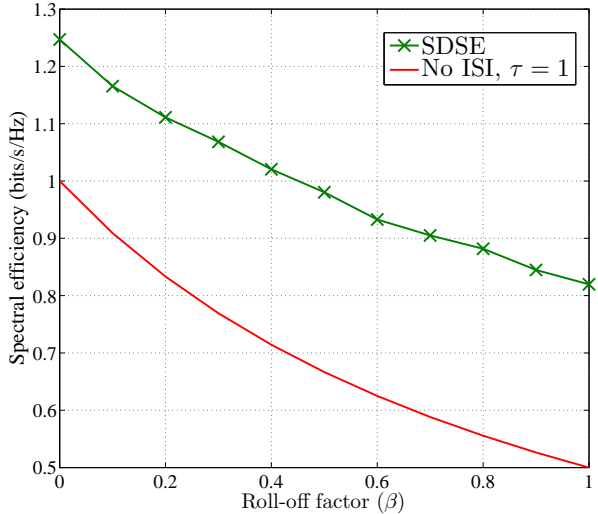}
	\caption{Spectral efficiency comparison of binary FTN signaling versus $\beta$ using the proposed SDSE and Nyquist signaling at BER = $10^{-4}$.}\label{fig:SE}
\end{figure}

Fig. \ref{fig:SE} compares the spectral efficiency  in bits/s/Hz of the proposed SDSE and Nyquist signaling  for $\beta \in [0, 1]$ and  BER =  $10^{-4}$. The value of $\tau$ of the proposed SDSE,  at each value of $\beta$, is chosen to be the smallest value that maintains BER = $10^{-4}$. As can be seen, the achieved spectral efficiency of FTN signaling  of the SDSE  is significantly higher than its counterpart of Nyquist signaling for the same value of the excess bandwidth, BER, and SNR. In particular, at $\beta = 0$ and $0.3$, the SDSE improves the spectral efficiency by approximately $25 \%$ and $38.85\%$, respectively, when compared to Nyquist signaling for the same BER and SNR. 
Additionally, it can be seen that the spectral efficiency of the proposed SDSE surpasses the highest spectral efficiency of Nyquist singling   (1 bit/s/Hz achieved at $\beta = 0$) for ranges of $\beta \in [0, 0.45]$.

\section{Conclusion} \label{sec:conc}

FTN signaling is a promising physical layer transmission technique that has the potential of improving the spectral efficiency if the appropriate mechanisms are in place in the receiver to handle the associated ISI. In this paper, we have proposed a novel sequence estimation techniques, namely SDSE, to estimate the transmit data symbols of binary FTN signaling. 
Simulation results show that the proposed SDSE  achieves the MLSE performance at reduced computational complexity. 
Results additionally revealed that up to $42.86 \%$ increase of the data rate at $\beta = 0.3$, compared to the Nyquist signaling, is possible without increasing either the bandwidth or the energy per symbol (i.e., power).  
Additionally, results showed that the spectral efficiency of the proposed SDSE for $\beta \in [0, 0.45]$  is higher than the maximum spectral efficiency of Nyquist signaling of 1 bit/s/Hz that is achieved at $\beta = 0$. 

%
\IEEEpeerreviewmaketitle




%



\ifCLASSOPTIONcaptionsoff
  \newpage
\fi



%

\bibliographystyle{IEEEtran}
\bibliography{IEEEabrv,C:/Users/e.bedeer/Sync/Dropbox/Publications/mybib_file}

%




\end{document}